\documentclass[prl,aps,superscriptaddress,amssymb,twocolumn]{revtex4-1}
\usepackage{graphicx}
\usepackage{amsmath}
\usepackage{latexsym}
\usepackage{dcolumn}
\usepackage{bm}
\usepackage{float}
\usepackage{graphicx,here}

\emergencystretch=\maxdimen
\begin{document}
\title{Experimental observation of node-line-like surface states in LaBi}

\author{Baojie Feng}
\thanks{Email: bjfeng@hiroshima-u.ac.jp}
\affiliation{Hiroshima Synchrotron Radiation Center, Hiroshima University, 2-313 Kagamiyama, Higashi-Hiroshima 739-0046, Japan}
\author{Jin Cao}
\affiliation{Beijing Key Laboratory of Nanophotonics and Ultrafine Optoelectronic Systems, School of Physics, Beijing Institute of Technology, Beijing 100081, China}
\author{Meng Yang}
\affiliation{Institute of Physics, Chinese Academy of Sciences, Beijing 100190, China}
\author{Ya Feng}
\affiliation{Ningbo Institute of Materials Technology and Engineering, Chinese Academy of Sciences, Ningbo 315201, China}
\affiliation{Hiroshima Synchrotron Radiation Center, Hiroshima University, 2-313 Kagamiyama, Higashi-Hiroshima 739-0046, Japan}
\author{Shilong Wu}
\affiliation{Graduate School of Science, Hiroshima University, 1-3-1 Kagamiyama, Higashi-Hiroshima 739-8526, Japan}
\author{Botao Fu}
\affiliation{Beijing Key Laboratory of Nanophotonics and Ultrafine Optoelectronic Systems, School of Physics, Beijing Institute of Technology, Beijing 100081, China}
\author{Masashi Arita}
\affiliation{Hiroshima Synchrotron Radiation Center, Hiroshima University, 2-313 Kagamiyama, Higashi-Hiroshima 739-0046, Japan}
\author{Koji Miyamoto}
\affiliation{Hiroshima Synchrotron Radiation Center, Hiroshima University, 2-313 Kagamiyama, Higashi-Hiroshima 739-0046, Japan}
\author{Shaolong He}
\affiliation{Ningbo Institute of Materials Technology and Engineering, Chinese Academy of Sciences, Ningbo 315201, China}
\author{Kenya Shimada}
\affiliation{Hiroshima Synchrotron Radiation Center, Hiroshima University, 2-313 Kagamiyama, Higashi-Hiroshima 739-0046, Japan}
\author{Youguo Shi}
\thanks{Email: ygshi@iphy.ac.cn}
\affiliation{Institute of Physics, Chinese Academy of Sciences, Beijing 100190, China}
\author{Taichi Okuda}
\affiliation{Hiroshima Synchrotron Radiation Center, Hiroshima University, 2-313 Kagamiyama, Higashi-Hiroshima 739-0046, Japan}
\author{Yugui Yao}
\thanks{Email: ygyao@bit.edu.cn}
\affiliation{Beijing Key Laboratory of Nanophotonics and Ultrafine Optoelectronic Systems, School of Physics, Beijing Institute of Technology, Beijing 100081, China}

\date{\today}

\begin{abstract}

In a Dirac nodal line semimetal, the bulk conduction and valence bands touch at extended lines in the Brillouin zone. To date, most of the theoretically predicted and experimentally discovered nodal lines derive from the bulk bands of two- and three-dimensional materials. Here, based on combined angle-resolved photoemission spectroscopy measurements and first-principles calculations, we report the discovery of node-line-like surface states on the (001) surface of LaBi. These bands derive from the topological surface states of LaBi and bridge the band gap opened by spin-orbit coupling and band inversion. Our first-principles calculations reveal that these ``nodal lines'' have a tiny gap, which is beyond typical experimental resolution. These results may provide important information to understand the extraordinary physical properties of LaBi, such as the extremely large magnetoresistance and resistivity plateau.

\end{abstract}

\maketitle

Topological semimetals represent a novel class of quantum materials whose conduction and valence bands touch at discrete points or extended lines\cite{Young2012,Wang2012,Liu2014,Xu2015,Burkov2011,Xu2011,Fang2015,KimY2015,YuR2015,Huh2016,Rhim2015,Wu2016,Schoop2016,Neupane2016,Bian2016prb,Bian2016nc,Takane2016,FengB2017}. Until now, three types of topological semimetals have been discovered: Dirac, Weyl and nodal line. In Dirac/Weyl semimetals, the energy-momentum dispersion is linear along all momentum directions, forming Dirac/Weyl cones in the proximity of the Fermi level. In nodal line semimetals, the crossing points form extended lines in the momentum space, i.e., the nodal lines\cite{Burkov2011,Xu2011,Fang2015,KimY2015,YuR2015,Huh2016,Rhim2015,Wu2016,Schoop2016,Neupane2016,Bian2016prb,Bian2016nc,Takane2016}. While the Dirac cones or nodal lines discovered in topological semimetals typically derive from the bulk bands, an interesting question is that whether they can exist as surface states of three-dimensional crystals. The nodal line surface states could manifest extraordinary properties that are distinct from other two-dimensional Dirac materials\cite{FengB2017,GeimAK2007,CastroNeto}. To date, the surface Dirac cones have already been extensively studied in topological insulators, such as the Bi$_2$Se$_3$ family materials\cite{Hasan2010,Qi2011,AndoY2013}, while theoretical and experimental studies on the surface nodal lines are still rare\cite{Wu2016}.

Recently, rare-earth monopnictide LaBi has been predicted to be a topological insulator based on first-principles calculations\cite{ZengM2015}. Moreover, magneto-transport measurements showed that LaBi hosts extremely large magnetoresistence (XMR)\cite{SunS2016,Tafti2016,Singha2017}, in analogy to some topological semimetals, such as Cd$_3$As$_2$\cite{LiangT2015,FengJ2015}, TaAs\cite{ShekharC2015}, and ZrSiS\cite{WangX2016,Singha2017PNAS}. These results have stimulated great research interest to search for the Dirac bands in LaBi. Recent angle-resolved photoemission spectroscopy (ARPES) measurements revealed that there are multiple surface Dirac cones on the (001) surface of LaBi\cite{NayakJ2016,WuY2016,NiuXH2016,LouR2017}. However, the details of the band structures are still controversial. J. Nayak et al.\cite{NayakJ2016} and X. H. Liu et al.\cite{NiuXH2016} reported two surface Dirac cones at the $\bar{X}$ point with an energy separation of 75 meV and 80 meV, respectively. In contrast, R. Lou et al.\cite{LouR2017} reported only one Dirac cone at the $\bar{X}$ point. Therefore, clarifying these controversies, which is essential to understand the extraordinary transport properties in LaBi, calls for further experimental and theoretical efforts.

In this Letter, we present the results of high-resolution ARPES measurements and first-principles calculations on the electronic structures of LaBi. We find that the surface states near the $\bar{X}$ points resemble Dirac nodal lines, bridging the band gap opened by spin-orbit coupling (SOC) and band inversion. In addition, our calculation results show that these node-line-like bands originate from the topological surface states (TSSs) of LaBi. These results may provide important information to understand the extraordinary physical properties in LaBi and could stimulate further research interest to search for the exotic physical properties associated with nodal line fermions.

\begin{figure}[t]
\includegraphics[width=8cm]{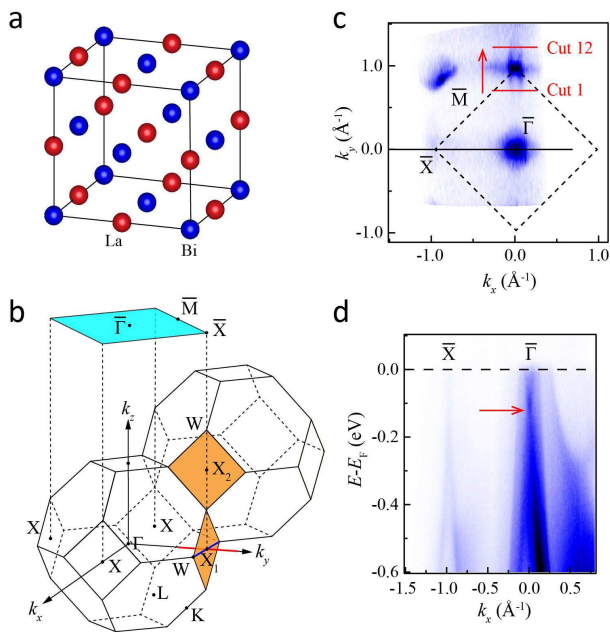}
\caption{(a) (Color online) Crystal structure of LaBi. Red and blue balls represent La and Bi atoms, respectively. (b) Bulk Brillouin zone and the (001)-projected surface Brillouin zone. The calculated band structures along the blue and red lines are shown in Figs. 4(a) and 4(b), respectively. (c) Photoemission intensity at the Fermi level of LaBi measured with 31-eV photons. Red lines indicate the momentum cuts along which the ARPES intensity plots in Fig. 2 were taken. (d) ARPES intensity plot along the black line in (c).}
\end{figure}

\begin{figure*}[htb]
\includegraphics[width=13cm]{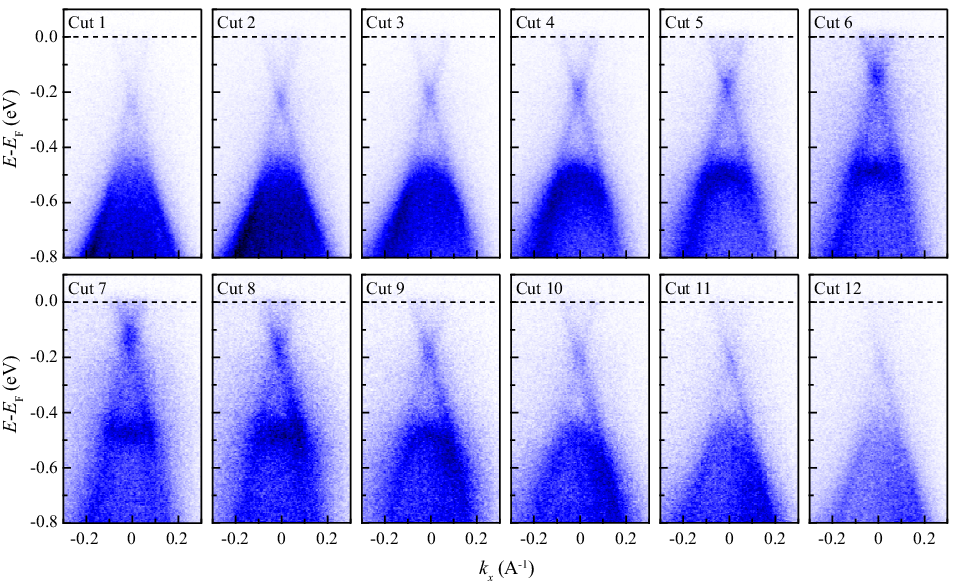}
\caption{(Color online) ARPES intensity plots near the $\bar{X}$ points. The data were taken along the red lines in Fig. 1(c). Cut 7 corresponds to the one that goes through the $\bar{X}$ point.}
\end{figure*}

LaBi crystalizes in a simple rocksalt-type structure (space group $Fm\bar{3}m$), as illustrated in Fig. 1(a). A schematic drawing of the three-dimensional BZ and the (001)-projected surface BZ is shown in Fig. 1(b). Each surface $\bar{X}$ point is projected from two inequivalent bulk $X$ points which are related by the $C_4$ rotation operator. As a result, the $\bar{X}$ points have $C_4$ rotational symmetry and thus are equivalent to each other.

The photoemission intensity of the Fermi surface of LaBi is shown in Fig. 1(c), which includes several Fermi pockets at the $\mathit{\bar{\Gamma}}$ and $\bar{X}$ points of the surface BZ. Figure 1(d) shows the measured band structures along the $\mathit{\bar{\Gamma}}$-$\bar{X}$ direction. At the $\mathit{\bar{\Gamma}}$ point, we observe a Dirac-like band with the crossing point located at approximately 0.1 eV below the Fermi level, as indicated by the red arrow. These results are consistent with previous ARPES measurements\cite{NayakJ2016,WuY2016,NiuXH2016,LouR2017}.

We then focus on the band structures near the $\bar{X}$ point. Figure 2 shows the band structures along a series of parallel cuts in the proximity of the $\bar{X}$ point, as indicated by the red lines in Fig. 1(c). One can see that there is a linear band crossing along each cut without any detectable gap feature. The Fermi velocity is approximately 7$\times$10$^5$ m$\cdot$s$^{-1}$, which is comparable to that in graphene. From cut 1 to cut 12, the crossing point gradually moves from $E\rm_B$=0.23 eV (cut 1) to $E\rm_B$=0.14 eV (cut 7), and then moves back to higher binding energies. These results indicate that there is a Dirac nodal line that goes through the $\bar{X}$ point. The nodal line is along the $\mathit{\bar{\Gamma}}$-$\bar{X}$ direction and has a weak dispersion. According to the $C_4$ rotational symmetry of the $\bar{X}$ point, there must be a perpendicular nodal line crossing the $\bar{X}$ point, but its intensity is much weaker because of the matrix element effect.

We realize that our experimental data are generally consistent with those in Ref.\cite{LouR2017}. In Ref.\cite{LouR2017}, the authors also observed a gapless band crossing at $\bar{X}$ with the crossing point located at $E\rm_B$=0.14 eV. However, the authors assigned these bands as a surface Dirac cone. From our experimental results, the linear crossing will not be gapped when we measure along neighboring cuts. Thus the bands near the $\bar{X}$ point form two perpendicular Dirac nodal lines, instead of a single Dirac cone.

In order to understand the intriguing nodal line features near the $\bar{X}$ point, we performed DFT calculations as implemented in the VASP package (see Supplementary Materials for details). All atomic positions and lattice parameters were fully relaxed before band structure calculations. Figure 3(a) shows the bulk bands along high-symmetry directions of the BZ, which is in general agreement with previous calculation results\cite{ZengM2015,GuoPJ2016}. Because of the SOC effects, the Bi $p$ band splits into a pair and the upper branch anti-crosses the La $d$ band along the $\mathit{\Gamma}$-$X$ direction. At the anti-crossing point, a small inverted gap ($\Delta\rm_{B}\approx$26 meV) develops due to band hybridization, as shown in Fig. 3(b). The inverted gap increases to approximately 0.4 eV at the $X$ point. For simplicity, we label the anti-crossing point $\mathit{\Sigma}$ hereafter.

The surface states of the (001) surface of LaBi were calculated using the Green's function method and the top five layers were included in the calculations (see Supplementary Materials for details). Figure 3(c) shows the calculation results along high symmetry directions of the surface BZ. Within the bulk band gap at the $\bar{X}$ point, there are linear ``crossing'' surface states that connect the conduction and valence bands, as indicated by the white arrow in Fig. 3(c). Actually, there is a tiny gap at the crossing point, which will be discussed later. The band crossing point is located at 0.3 eV below the Fermi level. The calculated band structures agree well with our experimental results except the different energy position of the crossing point. This discrepancy may originate from the different chemical potentials of our samples. To confirm whether the linearly crossing bands form a Dirac cone or a Dirac nodal line, we calculated the band structures along several parallel cuts near the $\bar{X}$ point, as shown in Figs. 3(d)-3(g). One can see that there is no increasing gap feature at the crossing point from Figs. 3(d) to 3(g), which confirms that the surface states near the $\bar{X}$ point form a nodal line. On the other hand, we realize that there is a relatively flat band that goes through the crossing point, as indicated by the blue arrow in Fig. 3(c). This band corresponds to the perpendicular nodal lines, but it is hardly observable in our experiments which may be caused by the matrix element effect. When we move away from the $\bar{X}$ point, this flat band splits into a pair: one moves upward while the other one moves downward, in agreement with the linear dispersion in the $k_y$ direction [Figs. 3(e)-3(g)].

Figures 3(h)-3(k) show constant energy contours near the $\bar{X}$ point at various binding energies. At 0.14 eV above the nodal line, the surface states form two ellipses that are perpendicular to each other [Fig. 3(h)], in agreement with the Fermi surface topology in Fig. 1(c). The ellipses become smaller at higher binding energies, and shrink to two perpendicular lines, i.e., the nodal lines [Fig. 3(j)]. Further increase of the binding energy leads to the separation of the bands, as shown in Fig. 3(k). These results fully support the nodal line shape of the surface states in the proximity of the $\bar{X}$ points.

\begin{figure*}[htb]
\includegraphics[width=17cm]{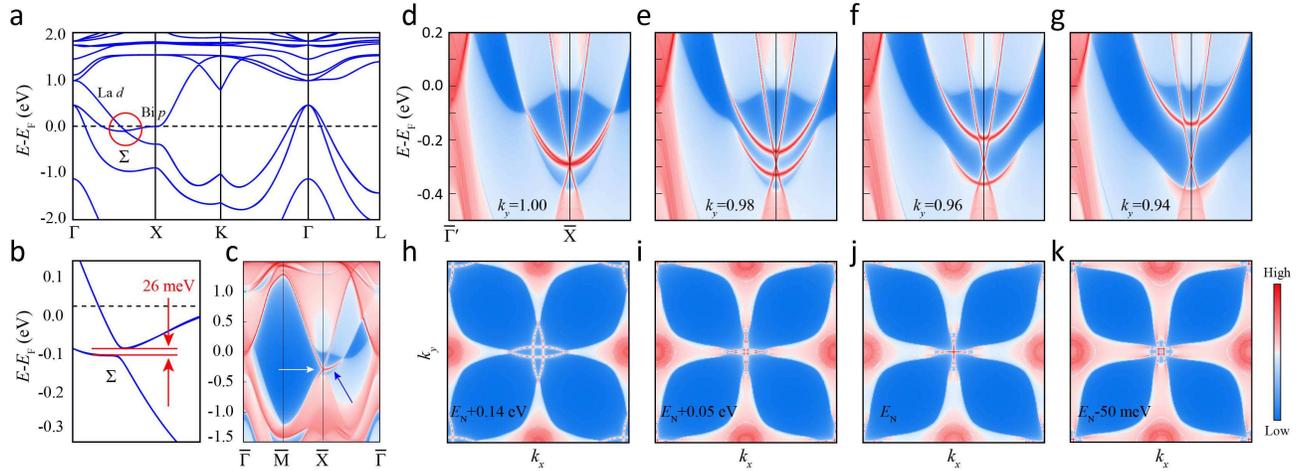}
\caption{(a) (Color online) Bulk bands of LaBi. The SOC effects were included in the calculations. (b) Zoom-in band structures in the red circles of (a). (c) Surface band structures of the (001) surface of LaBi. The white arrow indicates the linear band crossing; the blue arrow indicates a flat band that derives from the perpendicular nodal line. (d)-(g) Surface band structures near the $\bar{X}$ points at $k_y$=1.00, 0.98, 0.96, 0.94 $\pi/a$, respectively. $\mathit{\bar{\Gamma}}^\prime$ corresponds to the center of the second BZ. The definition of $k_x$ and $k_y$ were illustrated in Fig. 1(c). (h)-(k) Constant energy contours near the $\bar{X}$ point at different binding energies. $E\rm_N$ corresponds to the binding energy of the nodal line at the $\bar{X}$ point, i.e., -0.3 eV.}
\end{figure*}

To understand the origin of these node-line-like surface states, we performed further calculations. In the following, we will show that these bands originate from the superposition of topological surface states (TSSs) which derive from the anti-crossing feature of two inequivalent $X$ points (e.g. $X_1$ and $X_2$ in Fig. 1b). Because of the band inversion at the $X$ point, a topological surface state (TSS) will emerge, bridging the inverted gap of the bulk bands. As $\bar{X}$ is a time reversal invariant momentum, the TSS is doubly degenerate at the $\bar{X}$ point; the degeneracy will be lifted at momenta away from $\bar{X}$, forming conical band structures. Along the $W$-$X$-$W$ direction, the inverted gap has a minimum at the $X$ point, as shown in Fig. 4(a), and the surface bands disperse linearly [illustrated by blue lines in Fig. 4(d)]. When the momentum cut is rotated from $W$-$X$-$W$ [i.e. the blue line in Fig. 1(b)] to $\mathit{\Gamma}$-$X$-$\mathit{\Gamma}$ [i.e. the red line in Fig. 1(b)], the inverted bulk gap gradually decreases, and finally almost closes at $\mathit{\Sigma}$, as highlighted by the blue circle in Fig. 4(b). Meanwhile, the upper and lower branches of the TSS, which connect the conduction and valence bands respectively, gradually approach each other and nearly touch at $\mathit{\Sigma}$, as highlighted by the blue circles in Figs. 4(c) and 4(d). Therefore, the TSS in the proximity of $\bar{X}$ resembles a Dirac nodal line [Fig. 4(d)]. According to the $C_4$ rotational symmetry, there is a perpendicular ``nodal line'' which originates from band inversion near another $X$. Both ``nodal lines'' go through the $\bar{X}$ point at the same binding energy, resulting in a four-fold degeneracy at $\bar{X}$. However, the four-fold degeneracy will be lifted by band hybridization, forming two doubly degenerate crossing points, as shown in Fig. 4(e). Since the energy separation of the two crossing points is very small ($\Delta\rm_{SS}\approx$ 9 meV), the nodal line character is still preserved.

So far we have argued that the node-line-like surface states originate from the TSSs of LaBi, the SOC effects may play an important role on the evolution of these bands. When the strength of SOC is reduced, both $\Delta\rm_{SS}$ and $\Delta\rm_{B}$ gradually close, as shown in Figs. 4(h)-4(j). This result shows that reducing the strength of SOC will decrease the gap at the ``nodal lines''. On the other hand, reducing SOC will push the band anti-crossing point closer to $X$ because of a smaller splitting of the Bi $p$ band, as shown in Figs. 4(h)-4(j). As a result, the ``nodal lines'' will become shorter. In brief, our calculation results show that the gap size and length of the node-line-like surface states are tunable by changing the strength of SOC.

We notice that there is controversy in recent first-principles calculation results\cite{NayakJ2016,LouR2017}. While J. Nayak et al.\cite{NayakJ2016} claimed that there are two band crossings at $\bar{X}$, R. Lou et al.\cite{LouR2017} reported only one band crossing. Here, our calculation results have similarities to both of the previous results. On the one hand, the two band crossings in our results (Fig. 4c) are quite similar to the calculation results in Ref.\cite{NayakJ2016}, although our calculated energy separation of the crossing points are much smaller. On the other hand, if we neglect the tiny gap ($\Delta\rm_{SS}$), our calculated surface bands along the $\mathit{\bar{\Gamma}}$-$\bar{X}$ direction (Fig. 3c) will have only one band crossing at $\bar{X}$, similar to the results in Ref.\cite{LouR2017}. Despite these similarities, we want to emphasize that our results have significant advances. We revealed that the original TSSs that bridge the inverted gap are gapped out due to band hybridization. Moreover, we showed that the energy separation of the two band crossings is very small and surface states are approximately nodal lines.

\begin{figure*}[htp]
\includegraphics[width=17 cm]{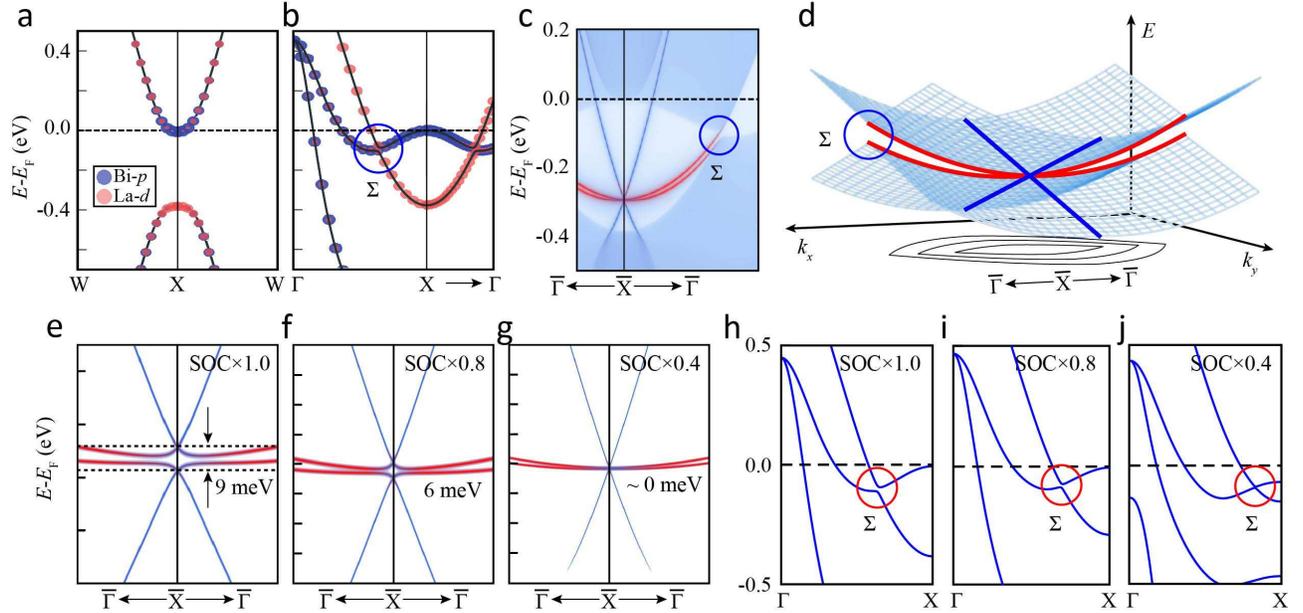}
\caption{(Color online) (a,b) Bulk band structures along the $W$-$X$-$W$ [cf. the blue line in Fig. 1(b)] and $\mathit{\Gamma}$-$X$-$\mathit{\Gamma}$ [cf. the red line in Fig. 1(b)] directions, respectively. (c) Zoom-in band structures in the proximity of the crossing point in Fig. 3(c). (d) Schematic drawing of the surface state from one of the bulk $X$ point. For clarity, the $C_4$-related surface state from an inequivalent bulk $X$ point is not shown here. The band dispersions along perpendicular directions is indicated by red (linear) and blue lines (quadratic). The constant energy contours are shown in the bottom. (e) Zoom-in band structures near the crossing point. The red and blue parts correspond to the contributions from two perpendicular ``nodal lines''. If band hybridization is neglected, the red (blue) part will show the quadratic (linear) dispersion indicated by the red (blue) lines in (d). (f,g) Surface band structures calculated after adjusting the strength of SOC to 80\% and 40\%, respectively. The energy separations of the two crossing points are indicated in the figures. (h-j) Bulk band structures along the $\mathit{\Gamma}$-$X$ direction after adjusting the strength of SOC. The red circles highlight the gap at the bulk anti-crossing point.}
\end{figure*}

Because of the finite experimental resolution, we did not observe the gap feature of the nodal lines in our ARPES measurements. Proving the existence of the tiny gap calls for higher resolution experiments. On the other hand, because of the negligibility of the gap, the physical properties related to the surface states of LaBi may be well explained by considering gapless nodal lines. Actually, a number of nodal line semimetals are not perfectly gapless, such as Cu$_3$PdN\cite{YuR2015} and ZrSiS\cite{Schoop2016,Neupane2016}. In addition, the intriguing spin texture of the nodal lines in LaBi, which originates from the topological surface states, may lead to a suppression of backscattering, despite the existence of a tiny gap. As a result, the extraordinary physical properties in LaBi, such as the XMR and resistivity plateau\cite{Tafti2016}, may be related to the node-line-like surface states, which remains to be understood.

In summary, we discovered node-line-like surface states on the (001) surface of LaBi based on combined ARPES measurements and first-principles calculations. These bands are located in the bulk band gap without detectable gap features. Our first-principles calculations reveal that these nodal lines derive from the TSSs of LaBi. These results provide important information on the electronic structures of LaBi, which may be crucial to understand its extraordinary physical properties. On the other hand, our results may stimulate further research interest to search for the novel physics of nodal line fermions.

\begin{acknowledgments}
B.F., J.C., M.Y. and Y.F. contributed equally to this work. The ARPES measurements were performed with the approval of the Proposal Assessing Committee of Hiroshima Synchrotron Radiation Center (Proposal Numbers: 17AG010 and 17AG011). We thank Professor X. J. Zhou for providing the Igor macro to analyse the ARPES data. This work is supported by the National Key R\&D Program of China (Grant No. 2016YFA0300600), the NSF of China (Grant Nos. 11734003, 11374338, 11774399, 11474330) and the Chinese Academy of Sciences (Grant Nos. XDB07020100, QYZDB-SSW-SLH043).
\end{acknowledgments}

\end{document}